\documentclass[prl,amssymb,twocolumn,tightenlines,showpacs]{revtex4}

\usepackage{graphicx}
\usepackage{amsmath}

\begin{document}

\title{Multiphoton path entanglement by non-local bunching}

\author{H.S. Eisenberg}
\affiliation{Department of Physics, University of California, Santa Barbara, California 93106, USA}

\author{J.F. Hodelin}
\affiliation{Department of Physics, University of California, Santa Barbara, California 93106, USA}

\author{G. Khoury}
\affiliation{Department of Physics, University of California, Santa Barbara, California 93106, USA}

\author{D. Bouwmeester}
\affiliation{Department of Physics, University of California, Santa Barbara, California 93106, USA}


\begin{abstract}
Multiphoton path entanglement is created without applying
post-selection, by manipulating the state of stimulated
parametric down-conversion. A specific measurement on one of
the two output spatial modes leads to the \textit{non-local
bunching} of the photons of the other mode, forming the
desired multiphoton path entangled state. We present
experimental results for the case of a heralded two-photon
path entangled state and show how to extend this scheme to
higher photon numbers.
\end{abstract}

\maketitle

Multiphoton path entangled states are superpositions of $n$
photons in one out of two or more paths. Such states can be
used to exceed the limitations imposed by the light
wavelength. One example is quantum photolithography where the
multiphoton interference of the different paths is used to
define details on a special photoresist film which are $1/n$
finer than the diffraction limit\cite{Boto,Fonseca}. The
photoresist should respond only to $n$ photons or more, which
is still an open challenge. Other uses are the enhancement of
the resolution of interferometric
measurements\cite{Holland,Jacobson,Kuzmich} and atomic
spectroscopy\cite{Bollinger,Leibfried}. In the context of
interferometry, path entangled states are a subset of a more
general group of usable photon-number correlated
states\cite{Holland,Silberhorn}.

Previously suggested methods to produce path entangled states
require either large non-linearities, non-unitary operations,
non-detection or include large statistical
bottlenecks\cite{Gerry,Dowling,Fiurasek,Hofmann}.
Non-detection can be replaced by post-selection, that actually
destroys the state. Furthermore, the schemes where $n>2$ rely
on the availability of various Fock states, which are
difficult to produce. Two photons from parametric
down-conversion can bunch to form a path entangled state
\cite{Rarity}, but this source is not expandable to larger
photon numbers. Recently, a state of three path entangled
photons was observed with post-selection through a
bottleneck\cite{Mitchell}.

In this Letter we present a way to create multiphoton path
entangled states without applying post-selection. The scheme
relies on two unique quantum-mechanical phenomena: bunching
(anti-bunching) of bosons (fermions) and non-locality. The
former reflects the discreteness and symmetries of the quantum
world. For example, it leads to the Hong-Ou-Mandel effect that
two indistinguishable photons entering a beam-splitter
simultaneously from both sides will always exit at the same
output port\cite{Hong}. The latter implies that two (or more)
distant particles can occupy a single quantum state and
possess correlations which no classical theory can
explain\cite{Bell}.

The scheme addressed in this Letter is based on the
manipulation of multiphoton entangled states that originate
from stimulated parametric down-conversion
(PDC)\cite{Lamas-Linares,Eisenberg}. By a specific measurement
of one of the PDC output spatial modes, the photons of the
other mode \textit{non-locally bunch} and form the desired
multiphoton path entangled state. It should be emphasized that
because no detection is needed at the second PDC mode, the
desired state is prepared by pre-selection. The detection in
the first mode is used as a heralding signal that announces
the creation of the path entangled state in the second mode.

We used non-collinear type-II parametric down-conversion with
spatial and spectral filtering to create the following
bi-partite state\cite{Braunstein}
\begin{subequations}\label{Psi}
\begin{eqnarray}
|\psi\rangle&=&\frac{1}{\cosh^{2}\tau}\sum_{n=0}^\infty\sqrt{n+1}\tanh^n\tau|\psi^-_n\rangle\,,\label{Psia}\\
|\psi^-_n\rangle&=&\frac{1}{\sqrt{n+1}}\sum_{m=0}^n(\textendash 1)^m|n\textendash m,m\rangle_a|m,n\textendash m\rangle_b\,,\label{Psib}
\end{eqnarray}
\end{subequations}
where $|m,n\rangle_i$ represents $m$ horizontally and $n$
vertically polarized photons in mode $i$. The magnitude of the
interaction parameter $\tau$ depends on the nonlinear
coefficient of the crystal, its length and the intensity of
the pump pulse. The state $\psi$, as well as its individual
terms $\psi_n^-$ of different photon-pair number $n$, are
invariant under mutual rotations of the polarization basis of
both spatial modes. The one-pair term ($n$=1) is the familiar
$\psi^-$ Bell state. We concentrate on the case when two
indistinguishable photon pairs are produced ($n$=2). It will
be subsequently shown how the presented method extends to
larger numbers of photons.

\begin{figure}[tbp]
\includegraphics[angle=-90,width=86mm]{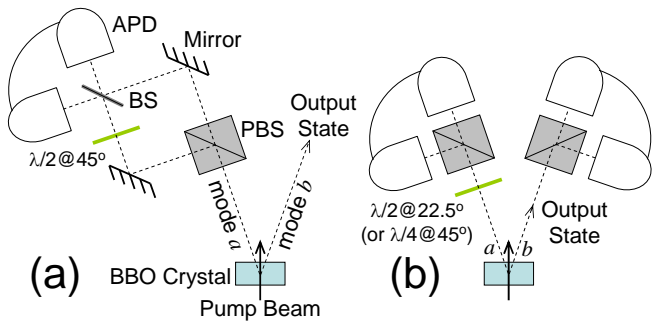}
\caption{\label{fig1}(\textbf{a}) A schematic setup for
non-local bunching of two photons. The polarized photons from
one of the spatial modes of PDC (mode $a$) are separated on a
polarization beam-splitter (PBS), the polarization of one arm
is rotated by a $\lambda/2$ waveplate at $45^\circ$ and the
two arms are combined on a beam-splitter (BS). Coincidence
detection is marked by connected single-photon avalanche
photo-diodes (APD). (\textbf{b}) A simpler but equivalent
scheme is to measure coincidence between the polarizations in
a rotated polarization basis ($\lambda/2$ at $22.5^\circ$ or
$\lambda/4$ at $45^\circ$).}
\end{figure}

The two-pair normalized term contains three equally weighted
elements:
\begin{equation}\label{Spin1Singlet}
|\psi^-_2\rangle=\frac{1}{\sqrt{3}}(|2,0\rangle_a|0,2\rangle_b-|1,1\rangle_a|1,1\rangle_b+|0,2\rangle_a|2,0\rangle_b)\,.
\end{equation}
The goal is to perform a measurement on mode $a$ that will
prepare mode $b$ in a path entangled state. To achieve this,
consider the setup presented in Fig. \ref{fig1}a. The middle
term of Eq. \ref{Spin1Singlet} has one horizontally and one
vertically polarized photon in mode $a$ as well as in mode
$b$. The photons in mode $a$ are separated by a polarizing
beam splitter (PBS). A $\lambda / 2$ waveplate in one of the
output arms of the PBS rotates the polarization in that arm to
the polarization of the other. The two photons from the middle
term, now indistinguishable in polarization, bunch on a 50/50
beam splitter (BS), therefore they can not give rise to a
coincidence detection between the two detectors\cite{Hong}.
Therefore, when such a coincidence is observed between the two
detectors in mode $a$, it could only have originated from the
first and third terms of Eq. \ref{Spin1Singlet}. The coherence
between the two terms is preserved by the coincidence
measurement, thus projecting the state of mode $b$ to
$|\psi_b\rangle=|0,2\rangle_b+e^{i\phi}|2,0\rangle_b$ (we drop
further normalization for simplicity). The phase $\phi$ is
determined by the difference between the lengths of the two
arms after the PBS in mode $a$. The coincidence detection
heralds the successful production of a path entangled state of
two photons in mode $b$. Entanglement in photon numbers is
created between two polarization modes rather than two paths.
A polarization beam-splitter and a $\lambda/2$ waveplate can
translate between the two representations.

Using the equivalence between the operation of beam-splitters
on two spatial modes and the operation of waveplates on two
polarization modes\cite{Shih}, it is possible to considerably
simplify the required coincidence measurement. As shown in
Fig. \ref{fig1}b, The two modes $a_h$ and $a_v$ bunch when the
polarization is rotated from the horizontal/vertical linear
polarization basis ($hv$) to either the plus/minus $45^\circ$
linear ($pm$) or right/left circular ($rl$). With the same
argument as above, it can be seen that coincidence at mode $a$
results in the desired state $\psi_b$ at mode $b$. Actually,
due to the rotational invariance of the state of Eq.
\ref{Spin1Singlet}, coincidence detection in mode $a$ at any
polarization state, implies bunching of mode $b$ in the other
two polarization bases. The difference between the bunched
states in the two bases is the sign between their two terms.

The non-local bunching result can be understood from another
point of view - starting from the detectors and propagating
backwards along the photon paths through the optical elements.
The detection operation is represented by annihilation
operators, e.g. $a_h$ for a detection of a horizontally
polarized photon in mode $a$. The two photon coincidence
detection operator in mode $a$ is the product $a_ha_v$. This
operator is transformed at a $\lambda/2$ waveplate to
$a_h^2-a_v^2$ and at a $\lambda/4$ waveplate to $a_h^2+a_v^2$.
Applying the transformed detection operator to mode $a$ of Eq.
\ref{Spin1Singlet} non-locally collapses mode $b$ to the
bunched state $\psi_b$ with an efficiency of 1/3:
\begin{eqnarray}\label{Psi2Bunched}
|\psi\rangle &=&(a_h^2+e^{i\theta}\cdot a^2_v)|\psi_2^-\rangle\\
\nonumber
&=&|0,0\rangle_a\otimes(|2,0\rangle_b+e^{i\theta}\cdot|0,2\rangle_b)\,,
\end{eqnarray}
where $\theta$ is a birefringent angle in the polarization
representation that equals 0 or $\pi$, depending on the choice
of measurement basis of mode $a$ ($rl$ or $pm$, respectively).

\begin{figure}[tbp]
\includegraphics[angle=0,width=86mm]{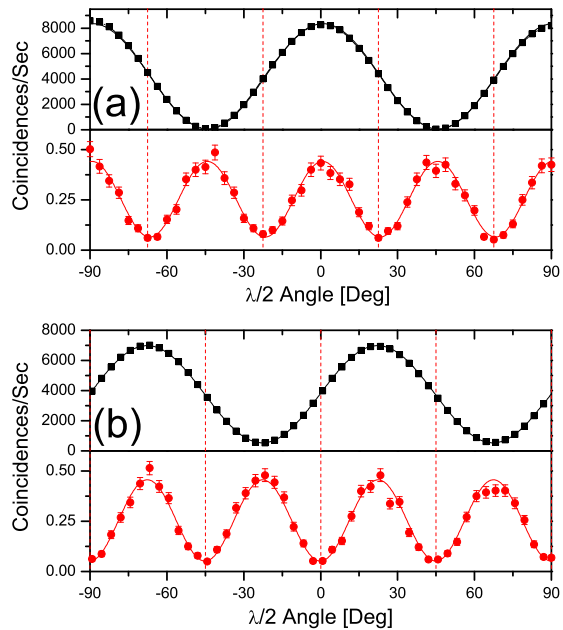}
\caption{\label{fig2} Two-fold (\textbf{squares}) and
four-fold (\textbf{circles}) visibilities and their fits
(solid lines). (\textbf{a}) Mode $a$ at the $hv$ basis and
mode $b$ is scanned between $hv$ and $pm$. (\textbf{b}) Mode
$a$ at the $pm$ basis and mode $b$ is scanned between $hv$ and
$pm$. Path entangled states are created at $\pm22.5^\circ$ and
$\pm67.5^\circ$ for \textbf{a} and at $0^\circ$,$\pm45^\circ$
and $\pm90^\circ$ for \textbf{b} (\textbf{dashed lines}).}
\end{figure}

In order to demonstrate non-local bunching, we down-converted
200\,fs pulses at 390\,nm in a BBO crystal with a double-pass
configuration\cite{Lamas-Linares}. From the measured rates,
the interaction parameter $\tau$ was evaluated to be about
0.1, thus the production ratio of three-to-two pairs was less
than $2\%$. The state of mode $b$ was analyzed by two
single-photon detectors $b_h$ and $b_v$ behind the
horizontally and vertically polarized output modes of a
polarization beam-splitter (Fig. \ref{fig1}b). To prove path
entanglement we first observed the absence of coincidences in
mode $b$, implying that photons travel in pairs, and then
examined the state coherence by interfering its two
components.

\begin{figure}[tbp]
\includegraphics[angle=0,width=86mm]{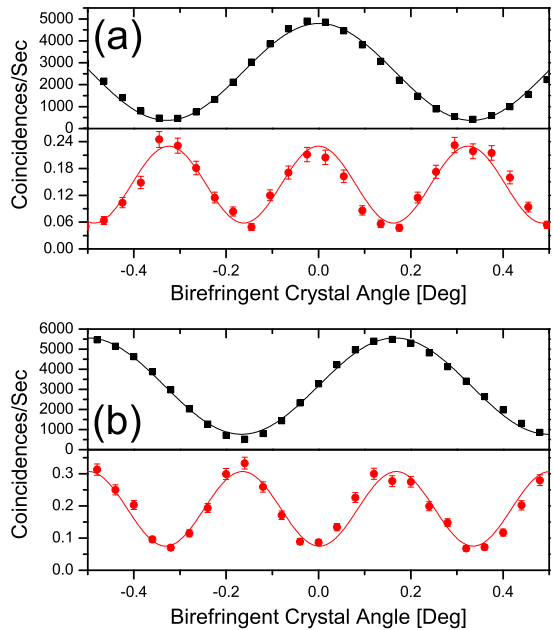}
\caption{\label{fig3} Two-fold (\textbf{squares}) and
four-fold (\textbf{circles}) coincidences as a function of the
birefringent phase. Fits are presented as solid lines.
(\textbf{a}) Mode $b$ is bunched at the $hv$ basis by
detecting mode $a$ in the $pm$ basis. (\textbf{b}) Mode $b$ is
bunched at the $hv$ basis by detecting mode $a$ in the $rl$
basis.}
\end{figure}

Visibility measurements were taken by fixing the polarization
basis of mode $a$ and recording various coincidences while
rotating the polarization basis of mode $b$. When the
polarization bases are different in the two modes, the
two-fold coincidence on mode $a$ bunches the photons in mode
$b$. The bunching prevents the $a_ha_vb_hb_v$ four-fold
coincidence and corresponds to the minima points
(\textbf{dashed lines}) in Fig \ref{fig2}. Tsujino \textit{et
al.}\cite{Tsujino} showed that the visibility of this
four-fold coincidence is related to the content $\alpha$ of
indistinguishable two photon-pairs, defined as
\begin{equation}\label{alpha}
|\psi\rangle=\sqrt{\alpha}|\psi_2^-\rangle+\sqrt{1-\alpha}|\psi_{1,\textrm{I}}^-\rangle\otimes|\psi_{1,\textrm{II}}^-\rangle\,,
\end{equation}
where Roman digits mark a distinguishing quantum number. In
their experiment they evaluated the content of the
indistinguishable state to be 37\%. Figure \ref{fig2} presents
visibility curves for two polarization settings. From the
measured four-fold visibility of $79\pm2\%$ we calculate
$\alpha$ to be $83\pm1\%$.

The visibility measurements indicate that a two-photon path
entangled state was produced in one of the two down-conversion
modes. In order to observe the presence of the two terms of
the state and their coherence, we interfered them on a
beam-splitter. We used again the analogy between the two
spatial modes of a beam-splitter and the two polarization
modes and interfered the $b_h$ and $b_v$ modes with a
$\lambda/2$ waveplate at $22.5^\circ$. Before the waveplate, a
phase $\theta_b$ between the two polarization modes was
introduced by tilting a birefringent crystal. As the
birefringent phase is scanned, $\theta$ of Eq. 3 varies as
$2\theta_b$ and the state behind the $\lambda/2$ waveplate
oscillates between $|2,0\rangle_b+|0,2\rangle_b$ and
$|1,1\rangle_b$ at twice the induced phase. These oscillations
were observed by detecting the four-fold coincidences of
$b_hb_v$ conditioned on $a_ha_v$. They are compared in Fig.
\ref{fig3} to the oscillations of the $a_hb_v$ two-fold
coincidence which follows $\theta_b$. The state of mode $b$
was bunched in the $hv$ basis, once by coincidence detection
of mode $a$ in the $pm$ basis (resulting in $\theta=\pi$) and
once in the $rl$ basis ($\theta=0$). Thus, at zero
birefringent phase the four-fold detection has a maximum in
the first case and a minimum in the second.

\begin{figure}[tbp]
\includegraphics[angle=-90,width=70mm]{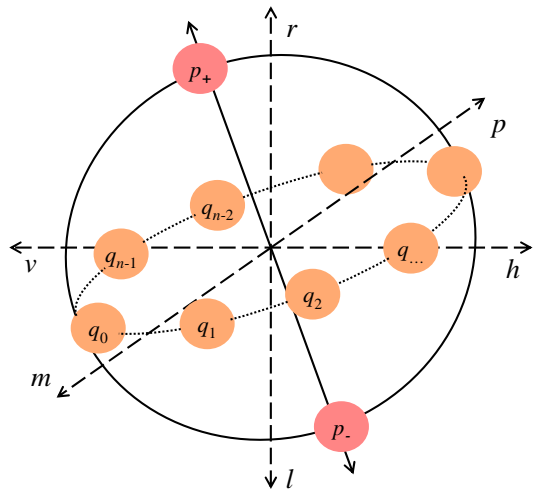}
\caption{\label{fig4} Poincar\'{e} sphere representation of
bunching of $n$ photons. The $n$ equidistant photons $q_m$
reside on a great circle and bunch at the $p_\pm$
polarizations.}
\end{figure}

In order to extend the scheme to higher photon numbers,
operations that bunch more detection operators should be used.
This is a generalization of the idea presented in Refs.
\cite{Dowling,Fiurasek,Hofmann}, but it does not require
non-detection or post-selection. In order to bunch $n$ photons
at a certain polarization basis $p$ (defined by an arbitrary
axis crossing the Poincar\'{e} sphere through its center, see
Fig. \ref{fig4}), one should combine $n$ polarized photons
$q_m$, residing equidistantly on the great circle whose plane
is perpendicular to that axis. The product of the $n$ linear
annihilation (or creation) operators $q_m$ is a different
representation for the bunched state in the $p$ basis:
\begin{eqnarray}\label{Poincare}
q_m&=&p_++e^{i\frac{2\pi m+\theta}{n}}\cdot p_-\,,\\
\nonumber \prod_{m=0}^{n-1}q_m&=&p_+^n-e^{i(n\pi+\theta)}\cdot p_-^n\,,
\end{eqnarray}
where $p_\pm$ are the annihilation operators at the two
polarizations defined by the axis. Notice how only two photons
$q_0$ and $q_1$ of orthogonal polarizations in an arbitrary
basis can have many great circles passing through them, thus
their bunching can be observed in any basis $p$ on the
equatorial between them.

\begin{figure}[tbp]
\includegraphics[angle=-90,width=86mm]{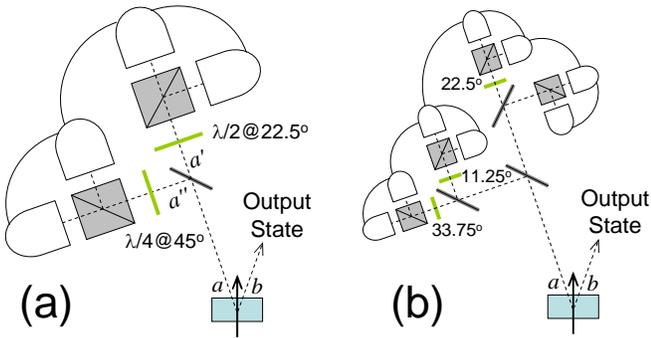}
\caption{\label{fig5}A schematic setup for extension of the
scheme for non-local bunching. Outputs are: (\textbf{a}) a
path entangled state of four photons in the $hv$ basis and
(\textbf{b}) eight photons in the $rl$ basis. All waveplates
in \textbf{b} are $\lambda/2$.}
\end{figure}

An example of a setup for four-photon path entanglement
generation is shown in Fig. \ref{fig5}a. The two output modes
of the beam-splitter are marked as $a'$ and $a''$. The
four-fold coincidence operator is $a_h'a'_va''_ha''_v$. It is
transformed by the two waveplates to
$({a'_h}^2-{a'_v}^2)({a''_h}^2+{a''_v}^2)$. The beam-splitter
combines the two modes operators to one $(a_h^4-a_v^4)$ with
an efficiency of ${4 \choose 2}=\frac{6}{16}$. Applying this
operator to the five terms of $\psi_4^-$ (Eq. \ref{Psib})
results in the heralded four-photon path entangled state with
an efficiency of 3/80:
\begin{eqnarray}\label{Psi4Bunched}
|\psi\rangle &=&(a_a^4-a^4_v)|\psi_4^-\rangle\\
\nonumber
&=&|0,0\rangle_a\otimes(|4,0\rangle_b-|0,4\rangle_b)\,.
\end{eqnarray}
Using a similar argument of rotational invariance, it is clear
that bunching of the photons in mode $b$ will occur in any
polarization basis as long as modes $a'$ and $a''$ are
detected in the other two complementary bases. A schematic for
the generation of an eight photon state is also shown (Fig.
\ref{fig5}b).

In conclusion, we demonstrated a scheme for a heralded source
of path entangled photon states by non-local bunching. The
photon resource is stimulated parametric down-conversion which
is relatively easy to produce compared to pure Fock states as
demanded by other proposals. The scheme is generally
expandable to higher photon numbers by using beam-splitters to
combine detection in different polarization bases. This
heralding detection signal can be used to open a polarization
insensitive switch in mode $b$, filtering out the lower photon
number content. The generation of a heralded two-photon path
entangled state was detected by observing interference at half
the photon wavelength.

The authors thank G.A. Durkin for fruitful discussions. This
research has been supported by NSF grants 0304678 and 0404440
and by DARPA MDA 972-01-1-0027 grant. H.S.E. acknowledges
support from the Hebrew University. J.F.H. thanks Lucent
Technologies CRFP for financial support. We acknowledge
support from Perkin-Elmer regarding the SPCM-AQR-13-FC single
photon counting modules.

\end{document}